\documentclass[twocolumn,superscriptaddress,prl,floatfix]{revtex4}

\usepackage{amsmath}
\usepackage{amssymb}
\usepackage{dsfont}
\usepackage{color}
\usepackage{graphicx}
\usepackage{epstopdf}
\usepackage{soul}
\usepackage{color}
\usepackage{braket}
\usepackage[normalem]{ulem}

\definecolor{SMblack}{RGB}{00,00,0}
\definecolor{Seba}{RGB}{00, 0, 00}
\usepackage[colorlinks=true,
            linkcolor=magenta,
            urlcolor=blue,
            citecolor=magenta]{hyperref}

\usepackage{lipsum}

\begin{document}

\title{Dissipatively Coupled Waveguide Networks for Coherent Diffusive Photonics} 

\author{Sebabrata~Mukherjee\footnote{These authors contributed equally to this work.} }
\email[Email: ]{s.mukherjee@hw.ac.uk}
\affiliation{Scottish Universities Physics Alliance (SUPA), Institute of Photonics and Quantum Sciences (IPaQS), School of Engineering $\&$ Physical Sciences, Heriot-Watt University, Edinburgh, EH14 4AS, United Kingdom}
\author{Dmitri Mogilevtsev$^*$}
\email[Email: ]{d.mogilevtsev@ifanbel.bas-net.by}
\affiliation{Institute of Physics, Belarus National Academy of Sciences, F. Skarina Ave. 68, Minsk 220072, Belarus}
\author{Gregory Ya. Slepyan}
\affiliation{Department of Physical Electronics, School of Electrical Engineering, Faculty of Engineering, Tel Aviv University, Tel Aviv 69978, Israel}
\author{Thomas H. Doherty}
\affiliation{School of Physics and Astronomy, University of St Andrews, North Haugh, St Andrews KY16 9SS, United Kingdom}
\author{Robert R.~Thomson}
\affiliation{Scottish Universities Physics Alliance (SUPA), Institute of Photonics and Quantum Sciences (IPaQS), School of Engineering $\&$ Physical Sciences, Heriot-Watt University, Edinburgh, EH14 4AS, United Kingdom}
\author{Natalia Korolkova}
\affiliation{School of Physics and Astronomy, University of St Andrews, North Haugh, St Andrews KY16 9SS, United Kingdom}
\date[]{Accepted for publication in Nature Communications}
\maketitle

{\bf A photonic circuit is generally described as a structure in which light propagates by unitary exchange and transfers reversibly between channels. In contrast, the term `diffusive' is more akin to a chaotic propagation in scattering media, where light is driven out of coherence towards a thermal mixture. Based on the dynamics of open quantum systems, the combination of these two opposites can result in novel techniques for coherent light control. The crucial feature of these photonic structures is dissipative coupling between modes, via an interaction with a common reservoir. Here, we demonstrate experimentally that such systems can perform optical equalisation to smooth multimode light, or act as a distributor, guiding it into selected channels. Quantum thermodynamically, these systems can act as catalytic coherent reservoirs by performing perfect non-Landauer erasure. For lattice structures, localised stationary states can be supported in the continuum, similar to compacton-like states in conventional flat band lattices.}


The engineering of dissipation to a common reservoir generates a vast array of novel structures for photonic application and quantum simulation. It has already been shown that the coupling of a number of quantum systems to the same reservoir gives rise to a decoherence-free subspace of Hilbert space~\cite{ekert}. Moreover, the evolution of an initial state towards this decoherence-free subspace is able to preserve and even create entanglement~\cite{zanardi,braun,benatti,lidar}. The careful engineering of loss can lead to coherence preservation~\cite{dav2001,man1,man2}, deterministic creation of non-classical states~\cite{zoll96,TFAb,parkins2003}, and serve as a tool for quantum computation~\cite{cirac2009,cirac2011}. Networks of dissipatively coupled systems
 were studied that can support topologically protected states~\cite{zollertop, Rudner, Zeuner}. Recently, optical setups were also used to study phenomena induced by engineered losses~\cite{Marandi, Pal, Tradonsky}.

Arrays of evanescently coupled optical waveguides is an excellent experimental platform to investigate a wide variety of semi-classical and quantum phenomena ranging from robust topological edge states~\cite{Rechtsman2013, Mukherjee2017} to quantum walks of correlated photons~\cite{Peruzzo2010}. Precise control in waveguide fabrication 
allows access to a desired Hamiltonian and the ability to probe the evolution of a specific initial state. Waveguide arrays with controllable loss and/or gain are also used to study various effects associated with non-Hermitian physics~\cite{Guo2009}.
In recent years, development of the femtosecond laser writing technique~\cite{davis} facilitated the fabrication of optical waveguides and waveguide based devices with three-dimensional geometry enabling the demonstration of intriguing phenomena known from condensed matter and quantum physics~\cite{Garanovich}.

Using the platform of integrated waveguide networks, here we propose that light can flow diffusively while remaining coherent and even entangled in a system of bosonic modes coupled to common reservoirs. In the experiment, performed using classical input states, we observed coherent diffusive equalisation in dissipatively coupled waveguide arrays.
Coherent diffusive light propagation opens new vistas for photonic applications, such as directional light distribution and diffusive coherence-preserving equalisation. In other words, we demonstrate that the aforementioned phenomena can be realised in the network of coupled integrated waveguides suggesting an exciting area in optical technologies, coherent diffusive photonics.\\

\begin{figure}[t!]
\centering
\includegraphics[width=1.0\linewidth]{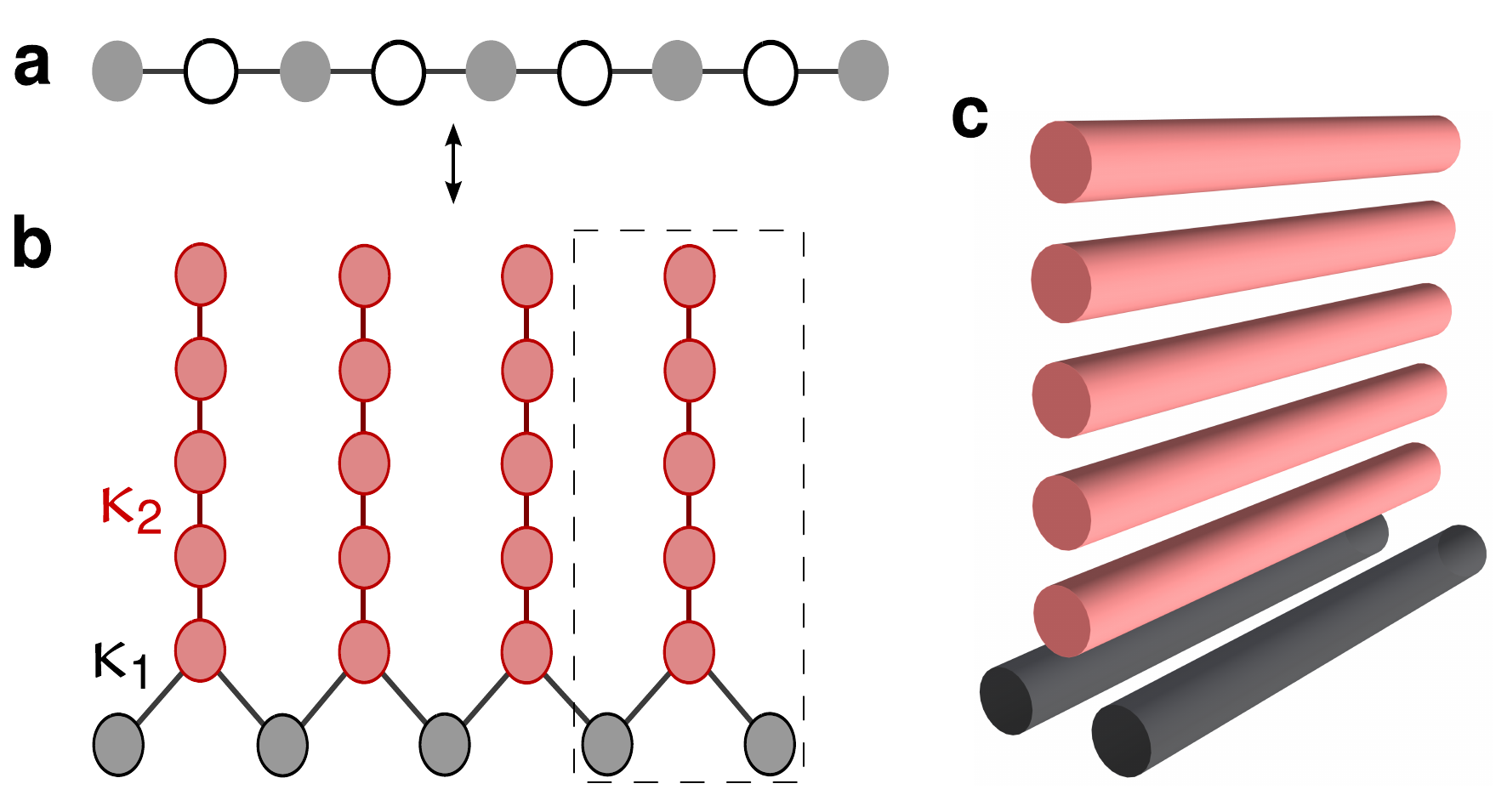}
\caption{{\bf One-dimensional diffusive photonic circuit. a,} Dissipatively coupled chain of the single mode waveguides. Every second waveguide (empty circles) is lossy and serves as a reservoir. The waveguides indicated by filled circles exhibit low loss and couple dissipatively via the auxiliary waveguides only. {\bf b,}  Experimental realization of the chain in {(a)} using an integrated optical circuit. The lossy sites are implemented using auxiliary arrangements of  coupled waveguides.  Here, $\kappa_1$ is the coupling between chain modes and lossy sites and  $\kappa_2$ is the coupling between the waveguides forming the reservoirs. {\bf c,} The three-dimensional geometry of the elementary diffusive circuit indicated by the dotted rectangle in (b).
\label{fig1}}
\end{figure}

{\bf Results}\\
{\bf Theoretical background.} 
The coherent diffusive photonic circuits considered in this article are described by the following generic quantum master equation:
\begin{eqnarray}
\frac{d}{dt}\rho= \sum\limits_{j=1}^N\gamma_{j}\left(2A_j\rho
A_j^{\dagger}-\rho A_j^{\dagger}A_j-A_j^{\dagger}A_j\rho\right),
 \label{chain1}
\end{eqnarray}
where $\rho(t)$ is the density matrix, $A_j$ denote the Lindblad operators for mode $j$ and $\gamma_j$ are the relaxation rates into corresponding reservoirs (see Supplementary Note 1 for more detail). In the experiments, we use femtosecond laser inscribed~\cite{davis} arrays of coupled optical waveguides, where the propagation of the light can mimic the time evolution described by specific Hamiltonians~\cite{Garanovich}. The relaxation rates $(\gamma_j)$ then describe the coherence diffusion rate between neighbouring waveguides. Coupling to common reservoirs is realised by mutually coupling each pair of waveguides to a linear arrangement of further waveguides~\cite{bigger}; see Fig.~\ref{fig1}.

Let us start with a simple example of 1D dissipatively coupled chain (DCC) with $A_j=a_j-a_{j+1}$, where $a_j$  ($a_j^{\dagger}$) is the bosonic annihilation (creation) operator, $a_j \vert \alpha_j \rangle= \alpha_j \vert \alpha_j \rangle$. Eq.(\ref{chain1}) can then be recast in terms of coherent amplitudes $\alpha_j$ (Supplementary Note 1):
\begin{eqnarray}
\frac{d}{dt}\alpha_k=-(\gamma_k+\gamma_{k-1})\alpha_k+
\gamma_k\alpha_{k+1}+\gamma_{k-1}\alpha_{k-1}.
 \label{cohr1}
\end{eqnarray}
Eq.~(\ref{cohr1}) formally coincide with the equations of a time-dependent classical random walk in one dimension, the discrete analogue of diffusion and heat transport dynamics. However, there are no classical probabilities in Eqs.~(\ref{chain1}, \ref{cohr1}), with the amplitudes $\alpha_j$ being complex. While the light flows diffusively, like heat, its coherence is maintained: off-diagonal elements in the Fock-state basis do not decay. For this to be the case, a fundamental role is played by the collective symmetrical superposition of all modes, characterised by a sum of modal operators:
\begin{eqnarray}
A_{\text{sum}}=\sum\limits_{j=1}^{N+1} \frac{a_j}{\sqrt{N+1}}. \label{Asum}
\end{eqnarray}

\begin{figure}[t!]
\centering
\includegraphics[width=1.0\linewidth]{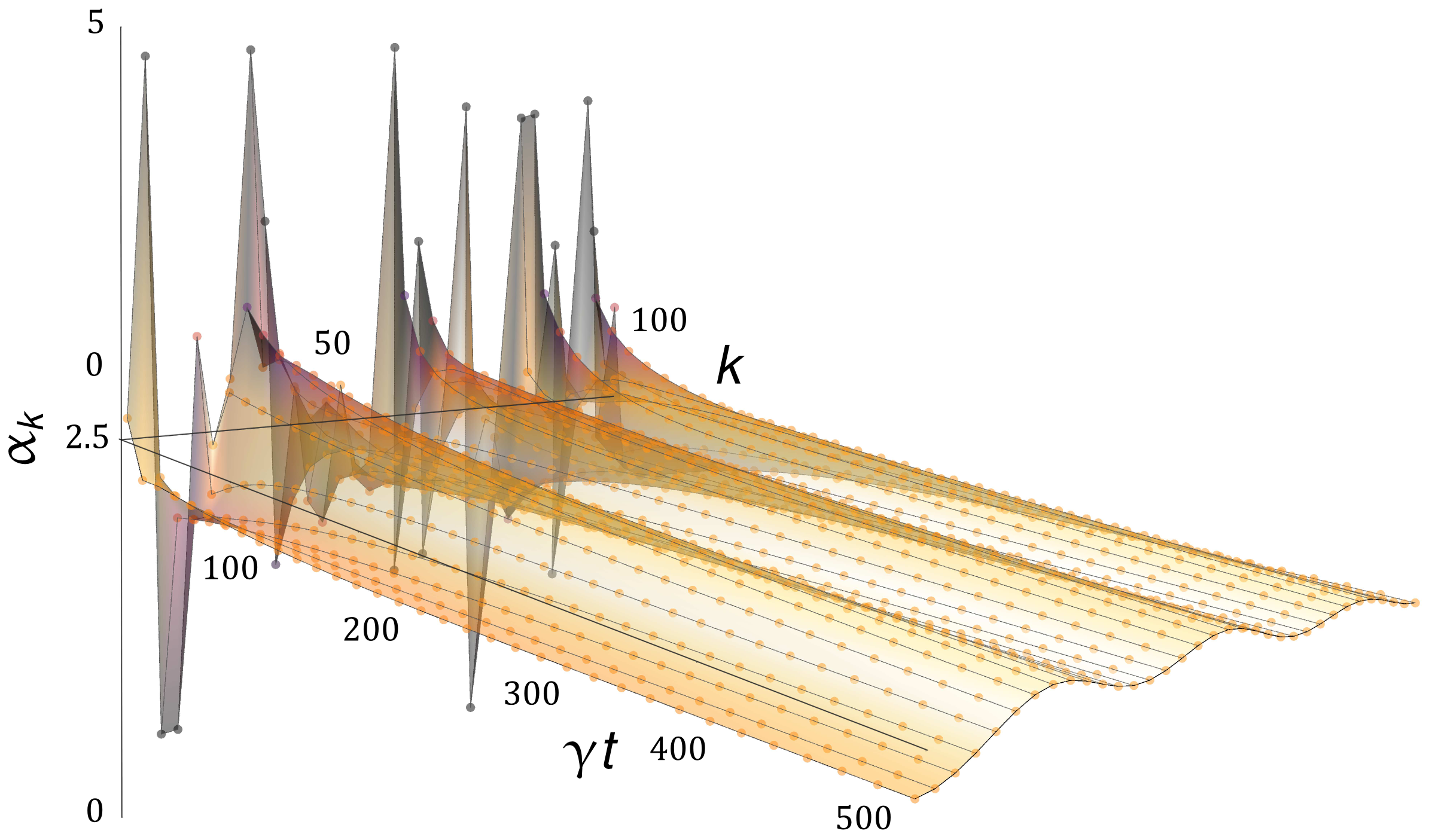}
\caption{{\bf Diffusive equalisation in a one-dimensional chain.} 
Simulation showing optical amplitude equalization for $k\!=\!100$ coherent states, propagating through a dissipatively coupled chain with equal coupling, $\gamma_j\!=\! \gamma$. The dynamics follow Eq.~(\ref{chain1}), where an initially random distribution of coherent state amplitudes are smoothed after a long effective time ($\gamma t$). Darker data points indicate a larger deviation from the mean amplitude, $\alpha\!=\!2.5$, emphasised by an equivalently coloured surface. Data points corresponding to each individual oscillator are connected.
 \label{fig2}}
\end{figure}

If a state is not symmetrical over all modes, it follows from Eqs.~(\ref{chain1}, \ref{cohr1}) that it will asymptotically decay to the vacuum state. Therefore, any state represented by a combination of operators $A_{\text{sum}}
$ and $A_{\text{sum}}^{\dagger}$
is conserved by the dynamics of Eq.~(\ref{Asum}). These states can be quite diverse in nature, from highly non-classical to Gibbs states (see Supplementary Note 3). Furthermore, a stationary state can also be entangled: for a single photon in the DCC, the state $A_{\text{sum}}^{\dagger}\prod\limits_{\forall j}|0\rangle_j$ is stationary (see ref.~\cite{our2015} for details).

Consider an initialization with all modal oscillators in coherent states. Eq.~(\ref{chain1}) shows that this will evolve into a product of coherent states with equal and averaged amplitudes, $\prod\limits_{\forall j}|\alpha_{\text{sum}}\rangle_j$ where ${\alpha_{\text{sum}}}=\sum\limits_{j}\alpha_j/(N+1)$. This feature of the diffusive, yet coherent 1D circuit, opens the possibility to realise an optical equaliser, suppressing both intensity and phase fluctuations in multimode fields. The equaliser performance is illustrated in Fig.~\ref{fig2}, where it is shown how the DCC can completely smooth any arbitrary zero-mean variations of the input.\\

\begin{figure*}
\centering
\includegraphics[width=0.81\linewidth]{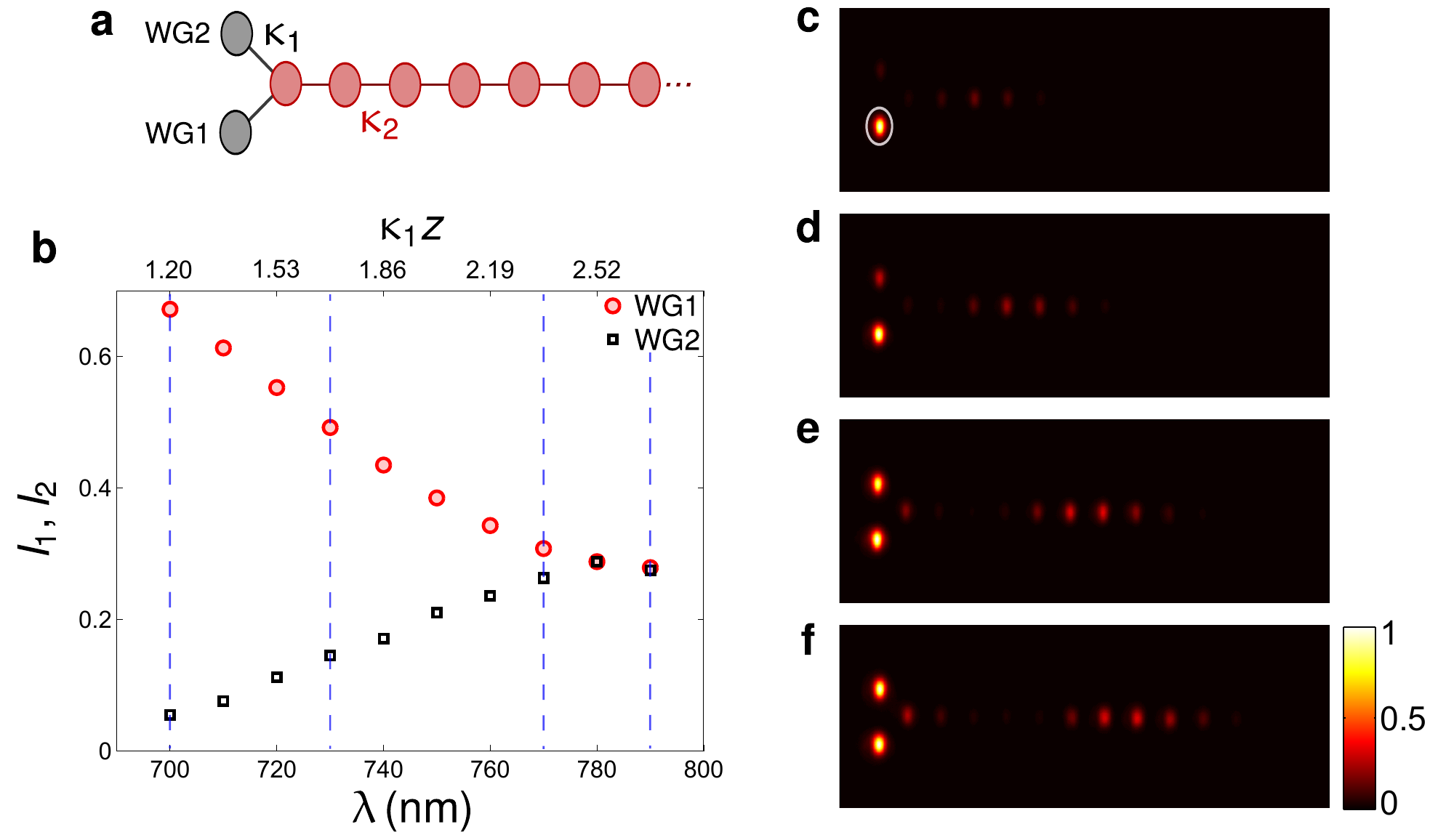}
\caption{{\bf Coherent diffusive equalisation in a dissipatively coupled waveguide pair.}  {\bf a,} Sketch illustrating the photonic implementation of two sites coupled via a reservoir (same as Fig.~\ref{fig1}~c). Here, WG1 and WG2 are two waveguides comprising the chain and the array of coupled waveguides, indicated by pink circles, acts as the reservoir. {\bf b,} The equalisation effect:~WG1 is excited initially with a coherent input and the light intensities $(I_{1,2})$ at WG 1, 2 are measured after a propagation of $z\!=\!30$ mm. The coherent diffusive evolution distributes the input light equally between WG1 and WG2. In this implementation, the effective propagation time  $\gamma t$ of Eq.~(\ref{chain1}) and Fig.~\ref{fig2} translates into propagation distance $z$ along the grey waveguides in Fig.~\ref{fig1}c. For a fixed sample length $z$, the dynamics can then be best assessed by monitoring the output intensity as a function of the wavelength as shown in the graph (see text and Supplementary Note 1, 2 for details). This effectively corresponds to changing $\gamma t\leftrightarrow\kappa_1z$. {\bf c-f,} Output intensity distributions for four different wavelengths [$700$~nm, $730$~nm, $770$~nm  and $790$~nm respectively], indicated by the vertical dashed lines in (b).
 \label{fig3}}
\end{figure*}
\begin{figure*}
\centering
\includegraphics[width=0.8\linewidth]{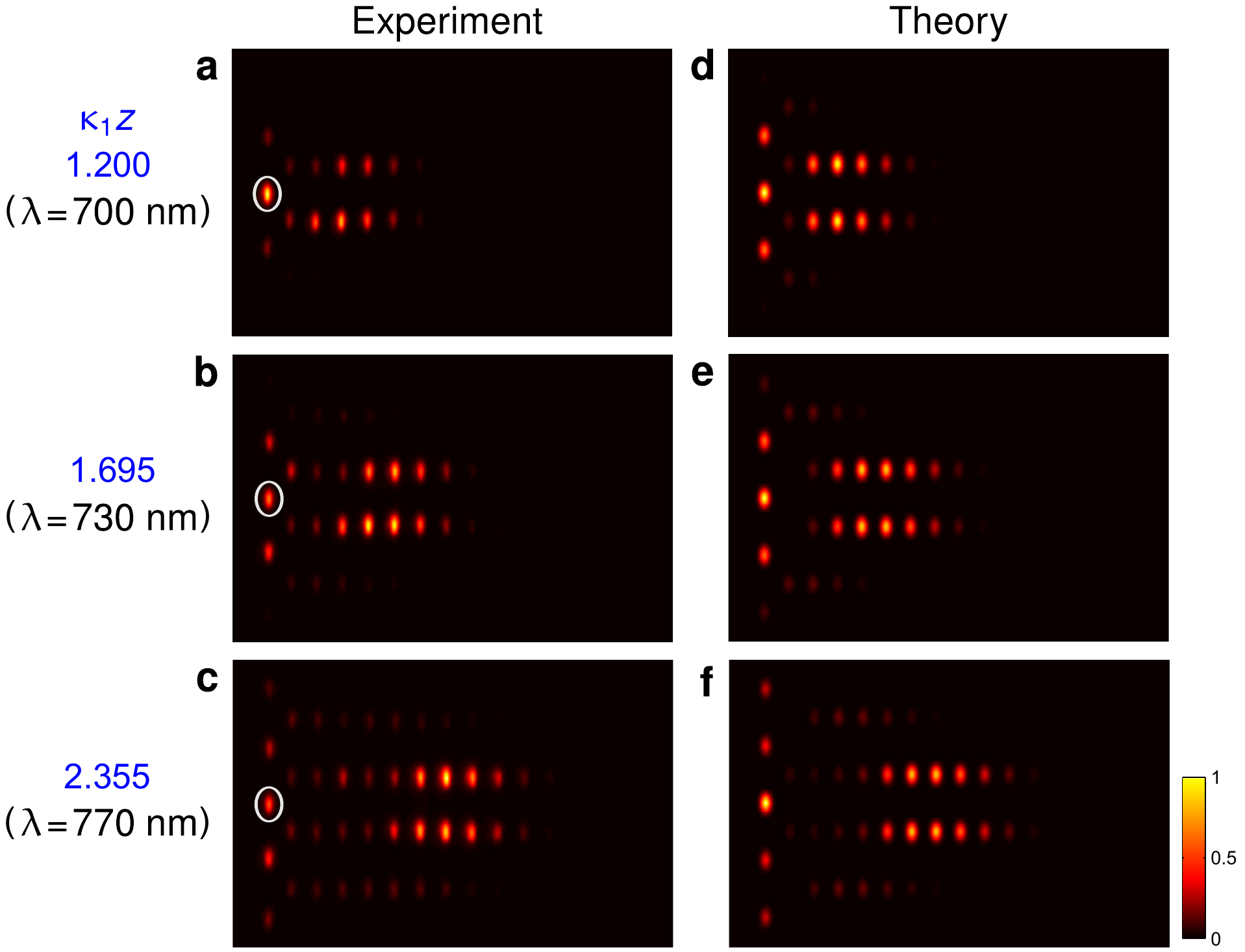}
\caption{{\bf Coherent diffusive equalisation in a dissipatively coupled waveguide array.}
 {\bf a-c,} Experimentally measured intensity distributions for three different values of $\gamma t\leftrightarrow \kappa_1z$. Here, the maximum propagation distance of the device is $z\!=\!30$~mm, $\kappa_1$ is tuned by varying the wavelength $(\lambda)$ of incident light, as in Fig.~\ref{fig3}. The central waveguide of the chain, indicated by the white circle, was excited at the input for all measurements. In this device, the waveguides in the chain are coupled via identical reservoirs, each containing twenty coupled waveguides. {\bf d-f,}  Numerically calculated intensity distributions corresponding to (a-c) respectively.
\label{fig4}}
\end{figure*}

{\bf Photonic implementation.}
In order to realise engineered dissipative coupling in integrated waveguides {in accordance with Eq.~\ref{chain1}}, one must be able to adiabatically eliminate lossy sites {(Fig.~1a)} from the system dynamics. This results in the fine-tuning of the evanescent coupling parameters $\kappa_{1,2}$, where the coupling between chain modes and lossy sites, $\kappa_1$, must be considerably smaller than intra--reservoir couplings, $\kappa_2$. For the particular design of Fig.~\ref{fig1}c, it  appears sufficient to have $\kappa_1/\kappa_2\!\approx\!0.5$ (see Supplementary Figure 1). The same ratio $\kappa_1/\kappa_2$ should hold for all $\gamma t$ (equivalently, $\kappa_1z$ in waveguide implementation, where $z$ is the propagation distance along the chain waveguides). The length of the DCC is not a prohibitive parameter and the collective behaviour, the coherent diffusive dynamics, can be established for merely two coupled bosonic modes. The effect of coherent optical equalisation can thus be readily achieved in the elementary circuit of Fig.~\ref{fig1}c. In the experiment, a 30-mm-long elementary circuit with 20 waveguides in the reservoir was fabricated (see Methods) and the output intensity distribution was measured as a function of the wavelength, $\lambda$, of incident light. It should be mentioned that both $\kappa_{1,2}$ vary linearly in the wavelength range of interest without affecting $\kappa_1/\kappa_2$ significantly, and hence, wavelength tuning enables us to observe the dynamics as the effective analogous time,  $\kappa_1(\lambda)z$, is tuned in this case (see Supplementary Note 2). Fig.~\ref{fig3} depicts the corresponding experimental results, clearly demonstrating equalisation of the input coherent signal. In the next step, we fabricated a chain of five waveguides, coupled via similar reservoirs, and demonstrate the coherent diffusive equalising. We excited the central waveguide of the chain at the input and measured the intensity distribution after a propagation of $z=30$~mm. Fig.~\ref{fig4} shows the output intensities for three different values of $\kappa_1z$. These experimental results are in good agreement with the numerically calculated output intensity distributions. Fig.~\ref{fig4} shows dynamically how the equalisation unfolds.\\

\begin{figure*}
\centering
\includegraphics[width=1\linewidth]{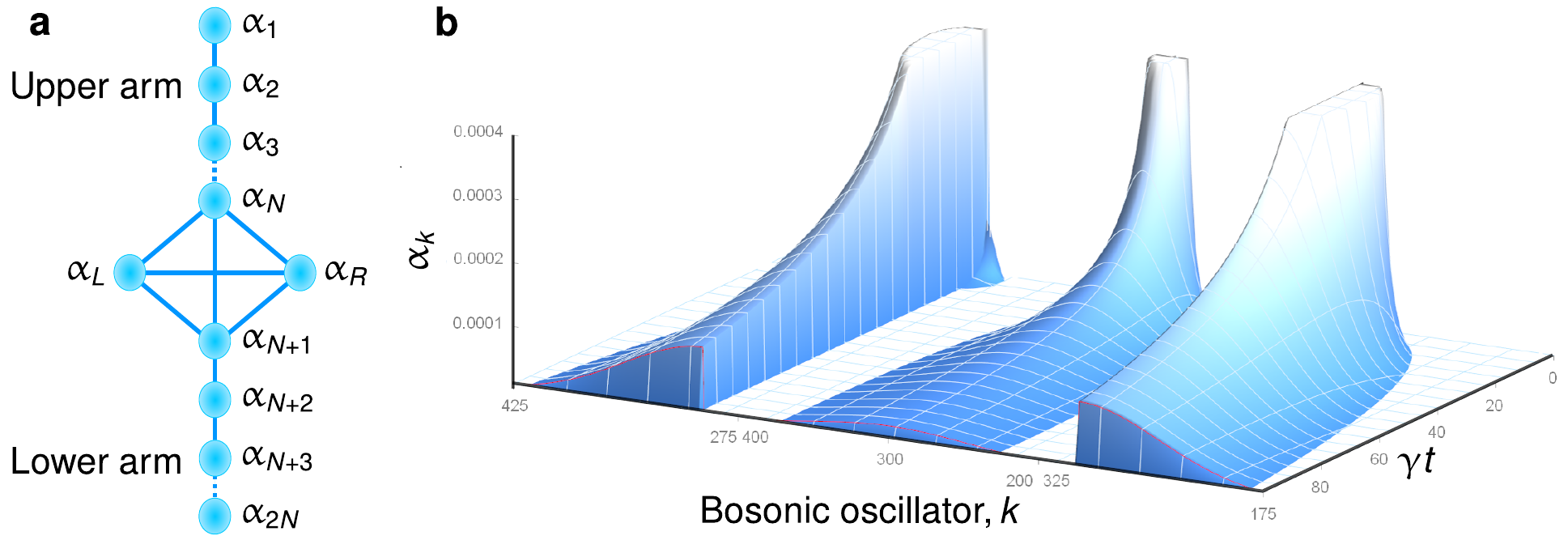}
\caption{{\bf Excitation distribution in a two-arm dissipative structure. a,} Illustration of two dissipatively coupled linear chains interacting via a pair of control modes. There are \textit{N} waveguides in each arm. {\bf b,} The coherent and guided transportation of light for three different states of the control modes. For these simulations, the number of waveguides in each arm is $N=300$, with only mode $\alpha_{N\!=\!300}$ being excited initially. In the leftmost distribution, the control modes, $\alpha_L$ and $\alpha_R$, have opposite phasing and light is guided into the lower arm. In the central scenario, $\alpha_L$ is excited and light equalises over the entire structure. In the rightmost case, the control modes are excited with equal amplitude and phase, confining light to the upper arm only.
\label{fig5}}
\end{figure*}

{\bf Two-dimensional diffusive circuits.}
When the linear arrangement of modes in the DCC is extended to further dimensions, a large vista of applications becomes accessible. These range from re-routing photonic devices to simulators of many-body quantum systems. Fig.~\ref{fig5}a outlines a photonic circuit for which the excitation of two control modes can dissipatively direct a coherent flow of light (Fig.~\ref{fig5}, b).  This ‘Quantum Distributor’ comprises two linear DCC, connected by mutual interaction to the pair of control modes (see Supplementary Note 4). Here, the control modes perform the distribution catalytically, their coherence being conserved.

Another simple DCC structure comprises two linear chains placed parallel with dissipative connections between each neighbouring mode (Supplementary Figure~2). This arrangement gives rise to the localization of signals which, unusually, is not born of defects (Supplementary Note~5). This is similar to the recently experimentally demonstrated lattice of unitarily coupled waveguides~\cite{erica,vic}. Alternative circuits are waveguides arranged as a honeycomb and square lattice (see Supplementary Figure 3 \& 4). The Lindblad operator for the honeycomb structure is  $L_j=\sum\limits_{k=1}^6(-1)^ka_{jk}$, where $j$ indexes hexagonal cells and $k$ numbers the modes in the cell. If each mode in a hexagonal cell has the same amplitude, the cell collectively constitutes a stationary, compacton-like state. These states satisfy $\langle L_j\rangle=0$, $\forall j$. It can be noted that they are robust with respect to additional losses in modes  neighbouring the cell. If there are some losses within the stationary cell itself, some non-vacuum states can still be supported (Supplementary Note~5). Moreover, coherence can spread diffusively in the lattice. Detailed discussion on the dissipative localisation in the diffusive square lattice can be found in Supplementary Note 5.\\

{\bf Quantum thermodynamical interpretation.}
The coherent diffusive dynamics of DCC also have an intriguing quantum thermodynamical interpretation. For a long DCC with identical initial coherent states of modes, the system will strongly equalise any fluctuations. We then dissipatively couple one further `signal' mode to this chain. This DCC will act as a reservoir, driving the signal mode towards some state independent of its initial excitation, asymptotically disentangled from the remainder of the chain. The state of the DCC after this interaction will belong to the same class of macro states as initially. Thus the long DCC chain is acting as a catalytic reservoir to the signal mode, and forthwith we use the term ‘reservoir’ to describe the arrangement.

Let us consider a DCC with $N+1\gg1$  oscillators in coherent states, each of amplitude $\alpha$, and the dissipatively connected signal mode having amplitude $a_0$. For an arbitrary initial state of the $a_0$ mode represented as $\rho_{a0}=\sum\limits_j p_j|\beta_j\rangle\langle\beta_j|_0$ \cite{ourprl2010}, the stationary state of the chain will be $\rho_{st}=\sum\limits_j p_j\prod\limits_{k=0}^{N+1}|{\bar\alpha}_j\rangle\langle{\bar\alpha}_j|_k$, with ${\bar\alpha_j}=\frac{1}{N+2}(\beta+(N+1)\alpha_j)$. For any finite set of $\beta_j$, the fidelity of the stationary state with the product of coherent states, $|\Phi\rangle=\prod\limits_{j=0}^{N+1}|\alpha\rangle_j$, tends to unity for large $N$. A sufficiently long DCC will, therefore, evolve any signal state into the coherent state initialised on the other oscillators in the chain. Hence, the DCC is indeed acting as a reservoir, washing away any information about the initial state. However, this clearly happens at a certain cost. The energy difference between the initial and asymptotic states of the chain and mode $a_0$ is given by:
\begin{eqnarray}
\Delta E=\frac{N+1}{N+2}\sum\limits_j p_j|\beta_j-\alpha|^2.
\label{energycost}
\end{eqnarray}
Note that in the limit of large $N$, the energy balance of the mode $a_0$ is the difference between the energies of the initial and the asymptotic state of the mode $a_0$: $\Delta E_0\approx \sum\limits_j p_j|\beta_j|^2-|\alpha|^2$. It can be equal to the energy loss of the whole mode $a_0$ plus DCC system, which holds for signal states satisfying $|\alpha|=\sum\limits_j p_j|\beta_j|\cos\{\arg(\beta_j)-\arg(\alpha)\}$.

Therefore, erasure of the state of mode $a_0$ can be performed without energy change of our reservoir. Such an action seems to contradict the famous Landauer's erasure principle: in order to erase information irreversibly, by an action of the environment, energy transfer into the environment needs to occur~\cite{landauer}. However, this principle was formulated for classical systems. The quantum Landauer's principle holds under different assumptions. These consist of the reservoir being a closed, Gibbs state system, which is entirely uncorrelated with the signal state. If the reservoir is not isolated from the environment, the applicability of the Landauer's principle is questionable~\cite{reeb}. Indeed, the use of an additional quantum system coupled to the reservoir allows the state of the signal to be erased without entropic change. The DCC is an example of the reservoir with such an additional quantum system.\\

{\bf Discussions}\\
In summary, we have illustrated intriguing possibilities for photonics that are generated by diffusive light propagation. The dissipative coupling of bosonic modes can allow light to flow like heat, whilst retaining coherence and even entanglement. A linear system of dissipatively coupled waveguides can act as an optical equaliser, smoothing fluctuations in amplitude and phase towards a common output. 
Any input state, classical or non-classical, will evolve into a completely symmetrised, correlated, state of the whole system.  This equalising action has been experimentally demonstrated with coherent input to an elementary photonic circuit (Fig.~\ref{fig3}) and for the chain of five waveguides (Fig.~\ref{fig4}). Further, we have outlined dissipative circuits which can catalytically direct the flow of light across multiple channels, or even support stationary lattice states without impurity. 

Generally, integrated waveguide networks lend themselves to applications in quantum information science~\cite{qucomm}, particularly in quantum communication. Quantum communication based on so-called {qumodes} (continuous variables optical quantum systems) has already proven feasible in terms of point-to-point transmission of quantum states (see ref.~\cite{Grosshans, Croal, Guenthner}) and this can be used in a number of applications such as quantum metrology and sensing, quantum cryptography and quantum signatures. Our coherent diffusive circuits are relevant as generic systems of qumodes propagating in integrated lossy networks where the loss mechanism provides quantum state engineering.

In the context of quantum thermodynamics, the DCC itself can be considered as a reservoir with non-trivial properties. Remarkably, the state of the reservoir can remain unchanged throughout the process of interaction with an external ``signal'' mode and allow non-Landauer erasure to be performed. Further, the optical equalisation and quantum evolution towards a stationary state can be used to study equilibration and thermalisation processes in quantum theory, one of the central problems in quantum thermodynamics ~\cite{equilibration}. In the future, we believe that diffusive photonic systems will find practical application both in studying the fundamental processes of structurally engineered open systems and in an array of integrated photonic technologies. Furthermore, the non-linear DCC can be engineered and implemented for producing and distributing non-Gaussian states~\cite{SM1}.\\

{\bf Methods.}  
The photonic devices formed by arrays of identical optical waveguides were fabricated using femtosecond laser writing technique~\cite{davis}. A $30$-mm-long borosilicate substrate (Corning Eagle$^{2000}$) was mounted on $x$-$y$-$z$ translation stages (ABL1000), and each waveguide was fabricated by translating the stages once through the focus of the fs laser pulses generated by an Yb-doped fibre laser (Menlo Systems, BlueCut; 350~fs, 500~kHz, and 1030~nm). The waveguide arrays were characterised using single-mode-fibre input coupling and free-space output coupling. To excite waveguides with a tunable wavelength of light, a photonic crystal fibre~\cite{Stone2008} was pumped using sub-picosecond laser pulses of $1064$~nm wavelength to generate a broadband supercontinuum. A tunable monochromator placed after the supercontinuum source was used to select narrow band ( $\!\sim\!3\!$~nm) light, which was coupled into an optical fibre (SMF-$600$). This fibre was then coupled to the desired waveguides. The output intensity distribution was observed using a CMOS camera (Thorlabs, DCC1545M).\\

{\bf Data availability}  Raw experimental data are available through Heriot-Watt University PURE research data management system (DOI: \href{https://doi.org/10.17861/15c1715e-a7c4-4bbf-beb4-91341f1c5ca0}{10.17861/15c1715e-a7c4-4bbf-beb4-91341f1c5ca0}).\\


\bigskip

{\bf Acknowledgements.} S.M.~and R.R.T.~sincerely thank the UK Science and Technology Facilities Council (STFC) for funding this work through ST/N000625/1. The authors acknowledge support from the EU projects FP7 People 2013 IRSES 612285 CANTOR (G.Ya.S.), Horizon-2020 H2020-MSCA-RISE-2014- 644076 CoExAN (G.Ya.S.), and SUPERTWIN id.686731 (D.M.),  the National Academy of Sciences of Belarus program ``Convergence" (D.M.). T.D.~and N.K.~acknowledge the support from the Scottish Universities Physics Alliance (SUPA) and the Engineering and Physical Sciences Research Council (EPSRC). 
The project was supported within the framework of the International Max Planck Partnership (IMPP) with Scottish Universities. D.M.~is thankful to Prof. A. Buchleitner and Dr.~V.~Shatokhin for useful discussions.

\bigskip
{\bf Author contributions.} D.M., G.Ya.S. and N.K. conceived the theory; S.M and R.R.T. devised the experiments; S.M.~designed, fabricated and characterised the photonic devices; T.D., D.M.~and S.M.~carried out theoretical calculations; D.M., S.M.~and N.K.~wrote the manuscript; N.K.~and R.R.T.~supervised the project; all authors discussed the paper.

\bigskip
{\bf Competing Interests.} The authors declare that they have no competing financial interests.


\newcommand{\beginsupplement}{%
        \setcounter{equation}{0}
       \renewcommand{\theequation}{S\arabic{equation}}%
        \setcounter{figure}{0}
        \renewcommand{\thefigure}{{\bf S\arabic{figure}}}%
     }
\beginsupplement
\onecolumngrid

\newpage
\begin{center}
\section*{\large Supplementary Information}
\end{center}
\begin{figure*}[h!]
\centering
\includegraphics[width=0.99\linewidth]{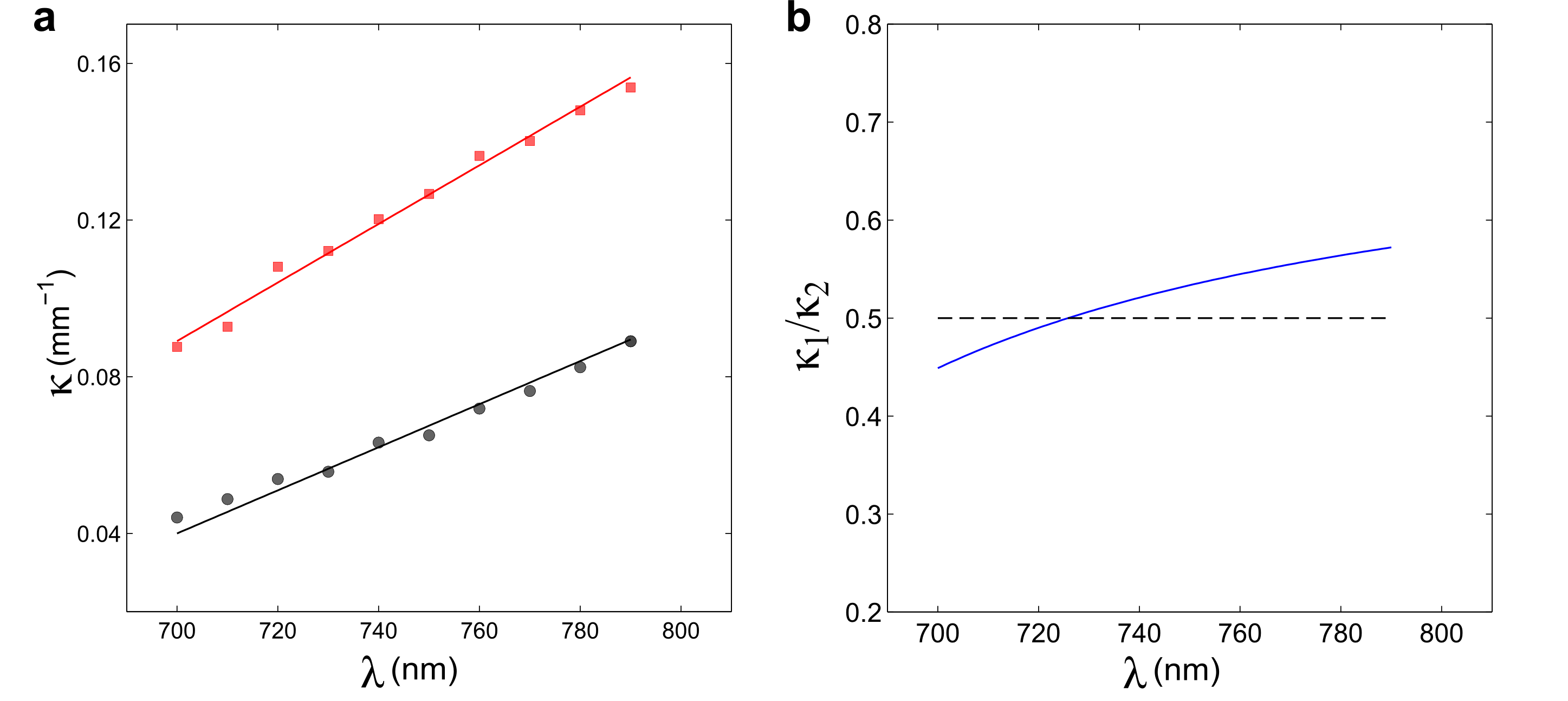}
\caption{{\bf {\color{SMblack}Measured variation of coupling strengths.}} {\bf a,} Variation of coupling constants ($\kappa_1 \rightarrow$ black and $\kappa_2 \rightarrow$ red) as a function of the wavelength of light. The solid lines are linear fits. {\bf b,} Variation of $\kappa_1/\kappa_2$ as a the function of wavelength of light. In the wavelength range of interest (i.e.~$700-790$~nm), this ratio remains very close to the desired value of $0.5$ (dashed line) with a maximal deviation of $\approx\pm0.05$.
\label{fig2-suppl}}
\end{figure*}
\begin{figure*}[h!]
\centering
\includegraphics[width=0.95\linewidth]{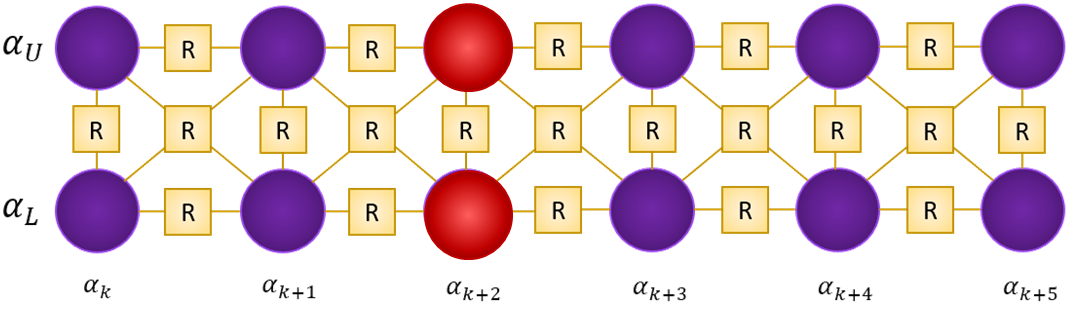}
\caption{{\bf {\color{SMblack} Schematic of a 2D diffusive photonic circuit (double chain).}}
{\color{SMblack}The circuit consists of two parallel dissipatively coupled chains. Here, the squares represent reservoirs (R), the circles are bosonic modes (the red circles indicate possible initial excitations).}
\label{fig3-suppl}}
\end{figure*}

\begin{figure*}[]
\centering
\includegraphics[width=0.45\linewidth]{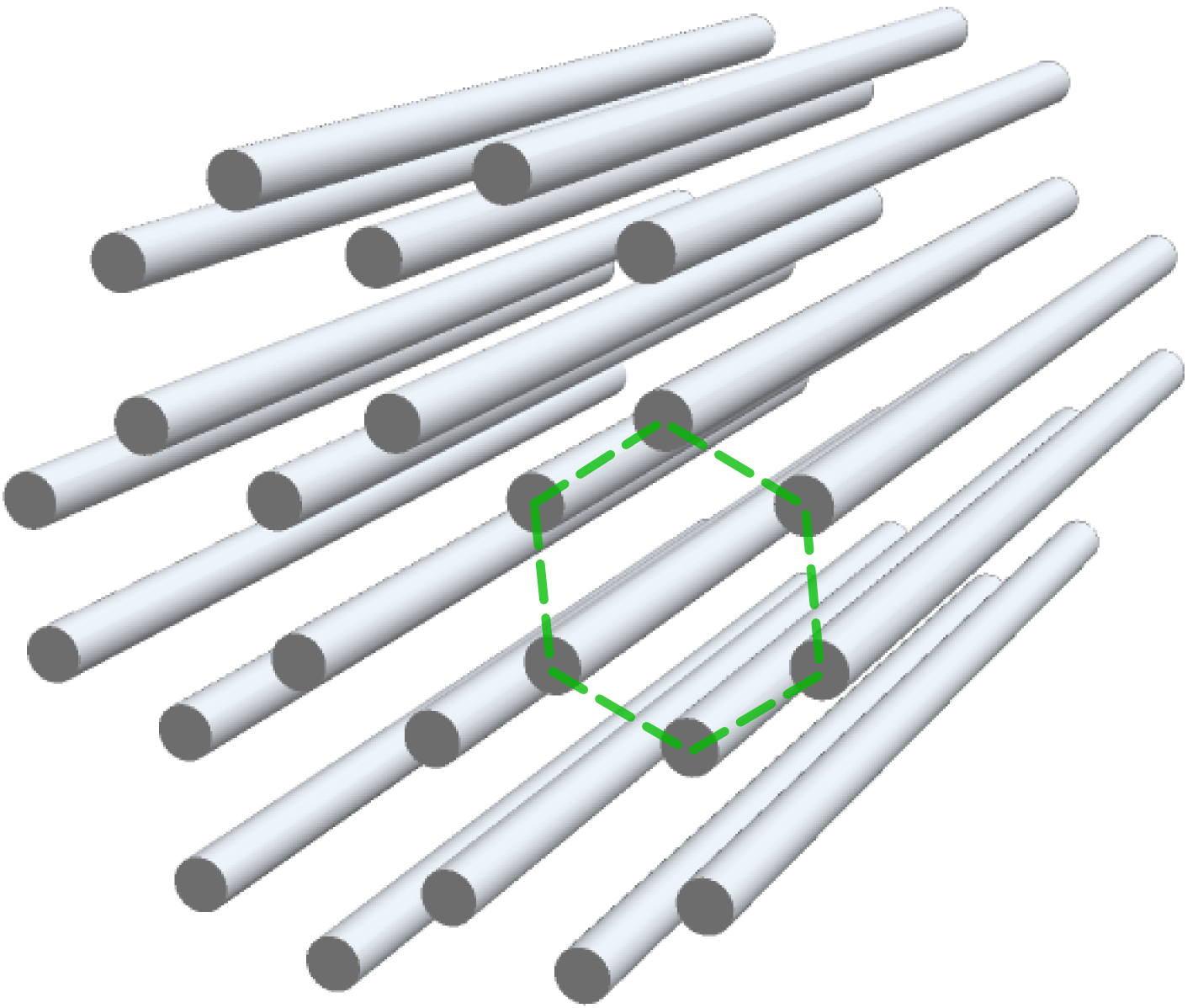}
\caption{{\bf {\color{SMblack} A diffusive photonic honeycomb lattice.}} 
{\color{SMblack} When all the bosonic modes in a hexagonal cell (indicated by the dashed line) have the same amplitude, the cell can support a stationary, compacton-like state.}
\label{fig4-suppl}}
\end{figure*}

\begin{figure*}[h!]
\centering
\includegraphics[width=0.8\linewidth]{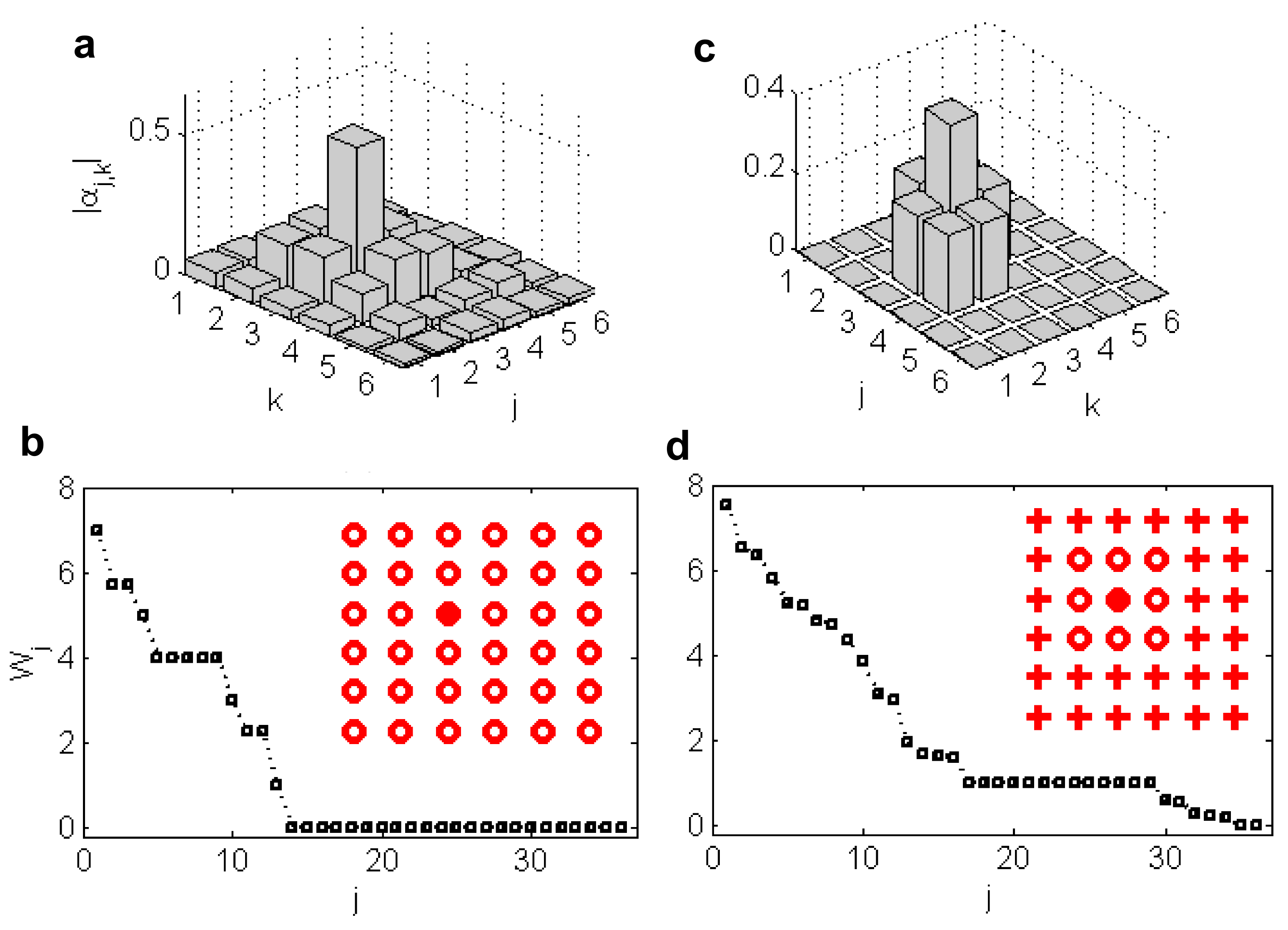}
\caption{{\bf {\color{SMblack} A diffusive square lattice.}} Stationary distributions of absolute values of modal amplitudes for {\color{SMblack}a} $6\times6$ square lattice {\color{SMblack}without (a) and with (c) additional losses at the sites indicated by red crosses in the inset of (d).
Eigenvalues (in the units of $\gamma$) of the systems [{\color{SMblack}Supplementary} Eq.~\ref{square eq}] without (b) and with (d) additional losses. Filled circles in the insets of (b,~d) denote the initial excitations.}
\label{sfig1}}
\end{figure*}

\clearpage
\section*{ Supplementary Note 1: Experimental model}

The coherent diffusive photonic circuits, considered in the main text, are described by the following generic quantum master equation:
\begin{eqnarray}
\frac{d}{dt}\rho= \sum\limits_{j=1}^N\gamma_{j}\left(2A_j\rho
A_j^{\dagger}-\rho A_j^{\dagger}A_j-A_j^{\dagger}A_j\rho\right),
 \label{chainS1}
\end{eqnarray}
where $\rho(t)$ is the density matrix  and $A_j$ denote the Lindblad operators for mode $j$. Quantities $\gamma_j$ are relaxation rates into corresponding reservoirs describing the coherence diffusion rate between neighbouring waveguides.
{\color{SMblack}Supplementary} Eq.~(\ref{chainS1}) with the Lindblad operator with $A_j=a_j-a_{j+1}$, where $a_j$  ($a_j^{\dagger}$) is the bosonic annihilation (creation) operator, follows from the usual model of the unitary coupled tight-binding chain of linear waveguides with every second waveguide being subjected to strong loss. The details of the derivation can be found, for example, in ref.~\cite{our2015}. 
This time evolution is modelled in the experiment by the arrays of coupled optical waveguides. The reservoirs are realised by mutually coupling each pair of waveguides to a linear chain of further waveguides as shown in Fig.~1 in the main text.

An interesting feature of the behaviour of light in these devices is the interchangeability between propagation distance and wavelength. The effective time of evolution $\gamma t$  can be altered both by changing the length of the waveguide block, or the wavelength of incident light. As the wavelength is tuned, $\kappa_1$ changes almost linearly to keep $\kappa_1/\kappa_2\!\approx\!0.5$, maintaining the correct character of dynamics.
Notice that the dependence of diffusion rates, $\gamma_j$, on time {\color{SMblack}changes} neither the diffusive character  of the dynamics nor the asymptotic state provided that always $\gamma_j(t)>0$.  

\section*{Supplementary Note 2: Measurement of evanescent coupling}
As mentioned in the main text, the control of evanescent coupling is crucial for the experimental realisation of the  diffusive equaliser. In {\color{SMblack}Supplementary} Fig.~\ref{fig2-suppl} we present the measured variation of $\kappa_{1,2}$ as a function of the wavelength of incident light, $\lambda$. We fabricated two types of directional couplers (each consisting of two evanescently coupled straight waveguides) which are the building blocks of the photonic circuits shown in {\color{SMblack}Fig.~1 (main text).} For the first type, where the two waveguides are at 45$^{\circ}$ angle, the coupling constant is $\kappa_1$ and that for the second type (consisting of two horizontally separated waveguides) is $\kappa_2$. Measuring the light intensities at the output of these 30-mm-long directional couplers, $\kappa_{1,2}(\lambda)$ were calculated~\cite{Szameit}.

It was observed that for these couplers, the ratio of $\kappa_{1,2}$ remains very close to the desired value of 0.5 with a maximum deviation of $\approx\pm0.05$.

\section*{Supplementary Note 3: Dynamics of the dissipatively coupled bosonic chain}

Due to {\color{SMblack}the} linearity of {\color{SMblack}Supplementary} Eq.~(\ref{chainS1}), the initial coherent states propagation through {\color{SMblack}DCC} remain coherent states at any time moment of dynamics described by {\color{SMblack}Supplementary} Eq.~(\ref{chainS1}).
Consider the Glauber $P$-function for the density matrix, $\rho(t)$, of the state describing the circuit:
\begin{eqnarray}
\rho(t)=\int d^2\vec{\alpha}P(\vec{\alpha},\vec{\alpha^{\ast}};t)|\vec{\alpha}\rangle\langle\vec{\alpha}|,
\end{eqnarray}
where $|\vec{\alpha}\rangle=\prod_j|{\alpha_j}\rangle$;
$|{\alpha_j}\rangle$ is the coherent state of the $j$-th mode of the circuit and the
amplitude $\alpha_j$ represents the $j$-th elements of the vector
$\vec{\alpha}$. For the {\color{SMblack}DCC} with $N+1$ modes and $j$th Lindblad operator represented as $A_j=a_{j}-a_{j+1}$, the
solution for the $P$-function is obtained from
the following Fokker-Planck equation: 
\begin{eqnarray}
\frac{\partial}{\partial t}P(\vec{\alpha},\vec{\alpha^{\ast}};t)=(\sum\limits_{j=1}^{N}\gamma_j(\frac{\partial}{\partial\alpha_j}\alpha_j-
\frac{\partial}{\partial\alpha_j}\alpha_{j+1}-\frac{\partial}{\partial\alpha_{j+1}}\alpha_{j}+\frac{\partial}{\partial\alpha_{j+1}}\alpha_{j+1})+{\rm h.c.}
)P(\vec{\alpha},\vec{\alpha^{\ast}};t)
\end{eqnarray}
Due to {\color{SMblack}the linearity of this equation,} the solution can be represented as
$P(\vec{\alpha},\vec{\alpha^{\ast}};t)=P(\vec{\alpha}(t),\vec{\alpha^{\ast}}(t))$,
where dynamics of amplitudes is described by Eq.~(2) of the main text. It is instructive to represent the initial state
in terms of discrete superposition of coherent state projectors~\cite{ourprl2010}:
\begin{eqnarray}
\rho(0)=\sum\limits_{\forall k}
p_k\prod\limits_{\forall j}|\alpha_{jk}\rangle\langle\alpha_{jk}|_j,
\label{datpat1}
\end{eqnarray}
where the index $k$ labels a set of amplitudes $\{\alpha_{k1},\alpha_{k2}\ldots \}$. The time-dependent Glauber function corresponding to the initial state [{\color{SMblack}Supplementary Eq.}~(\ref{datpat1})]
is given by {\color{SMblack}Supplementary} Eq.~(\ref{chainS1}) as
\begin{eqnarray}
P(\vec{\alpha},\vec{\alpha^{\ast}};t)=\sum\limits_{\forall k}
p_k\prod\limits_{\forall j}\delta(\alpha_j-\alpha_{jk}(t))\delta(\alpha_j^{\ast}-\alpha_{jk}^{\ast}(t)),
\end{eqnarray}
where amplitudes $\alpha_{jk}(t)$ for the {\color{SMblack}DCC} are defined from Eq.~(2) of the main text.

As follows from {\color{SMblack}Supplementary} Eq.~(\ref{chainS1}), any density matrix which is function of operators $A_{{\text{\color{SMblack}sum}}}=\sum\limits_{j=1}^{N+1} \frac{a_j}{\sqrt{N+1}}$, $A^{\dagger}_{{\text{\color{SMblack}sum}}}$, and
the vacuum, $\rho_{{\text{\color{SMblack}vac}}}=\prod\limits_{\forall j}|0\rangle\langle 0|_j$,
corresponds to a stationary state. These states can be of a quite
different nature. The stationary state can be just the pure
product of coherent states of individual modes with the same
amplitude:
\begin{eqnarray}
\rho_{{\text{st}}}=|\Phi_{{\text{st}}}\rangle\langle\Phi_{{\text{st}}}|, \quad
|\Phi_{{\text{st}}}\rangle=\prod\limits_{\forall j}|\alpha\rangle_j.
 \label{stationarycoh}
\end{eqnarray}
However, it can also be quite exotic, for example, it can be a Schr\"{o}dinger-cat
entangled state with $|\Phi_{{\text{st}}}\rangle\propto \sum\limits_{k=1}^K
w_k\prod\limits_{\forall j}|\alpha_k\rangle_j$, where $K$ is the
number of different components in our cat-state and $w_k$ are
scalar weights. The Gibbs state
\begin{equation}
\rho_{{\text{st}}}=\frac{{\exp\{-\beta
A_{{\text{\color{SMblack}sum}}}^{\dagger}A_{{\text{\color{SMblack}sum}}}\}}}{{{\rm Tr}\{\exp\{-\beta
A_{{\text{\color{SMblack}sum}}}^{\dagger}A_{{\text{\color{SMblack}sum}}}\}\}}}
\label{gibbs}
\end{equation}
also belongs to the stationary states of the system. This state
has maximal entropy for the given sum of the
second-order coherences, $\langle a^{\dagger}_k a_l\rangle$ (which
is also conserved by the dynamics). As was already mentioned, the stationary state can also be maximally entangled.

The {\color{SMblack}DCC} evolves toward {\color{SMblack} a} stationary state in a quite remarkable way. The initial state of the {\color{SMblack}DCC} with $N+1$ modes corresponding to the coherent state of all the
chain modes, $|\Phi_{0}\rangle=\prod\limits_{\forall
j}|\alpha_j\rangle_j$,  evolves to the product of coherent states with
equal amplitudes,
$|\Phi_{t\rightarrow\infty}\rangle=\prod\limits_{\forall j}|
\alpha_{{\text{\color{SMblack}sum}}}\rangle_j$, where the amplitude is the averaged sum of all
the amplitudes, ${\alpha_{{\text{\color{SMblack}sum}}}}=\sum\limits_{j}\alpha_j/(N+1)$.
Then an arbitrary initial state of the {\color{SMblack}DCC} [{\color{SMblack}Supplementary} Eq.~(\ref{datpat1})]
will be asymptotically reduced to the following form:
\begin{eqnarray}
\rho_{{\text{\color{SMblack}st}}}=\sum\limits_{\forall k} {p}_{k}\prod\limits_{j=1}^{N+1}|{\bar\alpha}_{k}\rangle\langle{\bar\alpha}_{k}|_j
 \label{ppositivefinal}
\end{eqnarray}
with
${\bar\alpha}_k=\frac{1}{N+1}\sum\limits_{j=1}^{N+1}\alpha_{jk}$. Actually, the
{\color{SMblack}DCC} drives the initial state to the symmetrical state over all the
modes.   Note, that the smoothing action of {\color{SMblack}DCC} is preserved even for the case of different decay rates, $\gamma_j\neq 0$. Stationary states do not depend on
them.

\section*{Supplementary Note 4: {\color{SMblack}Two-arm} distributor structure}
The prerequisite of the distributing action considered here is the existence of several localised stationary states of the structure described by the master equation, {\color{SMblack}Supplementary} Eq.~(\ref{chainS1}). For {\color{SMblack}the sake of} simplicity, we consider here pure stationary states.  We call the state $|\chi^{{\text{\color{SMblack}loc}}}\rangle$ ``localised" if exists some subset, $\{m\}$, of $M\!<\!K$ systems of our dissipatively coupled photonic circuit such that $\sum\limits_{k\in\{m\}}\!\langle\chi^{{\text{\color{SMblack}loc}}}|a_k^{\dagger}a_k|\chi^{{\text{\color{SMblack}loc}}}\rangle \!>\!0$, whereas for systems out of the subset $\{m\}$ we have
$\sum\limits_{k\notin\{m\}}\!\langle\chi^{{\text{\color{SMblack}loc}}}|a_k^{\dagger}a_k|\chi^{{\text{\color{SMblack}loc}}}\rangle\!=\!0$. The most simple and obvious distributing action would be possible if the initial state of the structure, $\rho_0$, is orthogonal to
some localised stationary state, {\color{SMblack} $\langle\chi^{{\text{loc}}}_j|\rho_0|\chi^{{\text{loc}}}_j\rangle=0$. }
Then the part of the
structure corresponding to subset $\{m_j\}$ will not be excited in
the process of dynamics described by {\color{SMblack}Supplementary} Eq.~(\ref{chainS1}). For such a distributor to be non-trivial,  sets corresponding to different localised states,  $\{m_j\}$, have to  be partially overlapping. {\color{SMblack} The distributor} can be realised even in the case when the localised stationary states corresponding to different parts of the structure are not mutually orthogonal.

Let us illustrate our consideration with the example of the slightly modified  {\color{SMblack}DCC}. For the structure depicted in Fig.~5a of the main text, modes in the arms are coupled 
pairwise, $A_j=a_j-a_{j+1}$ for $j=1\ldots N-1, N+2\ldots 2N$. 
For the central controlling node $L_N=a_N-{\color{SMblack}a_R}+a_L-a_{N+1}$. From the
master equation, {\color{SMblack}Supplementary} Eq.~(\ref{chainS1}), the equation similar to Eq.~(2) of the main text can be obtained for each arm. For four modes of the central node the equations are as follows: 
\begin{gather}
\frac{d}{dt}\alpha_N=-(\gamma_N+\gamma_{N-1})\alpha_N+
\gamma_{N-1}{\color{Seba}\alpha_{N-1}}+ \gamma_N(\alpha_{N+1}+\alpha_{H}-\alpha_{L}),\\
 \frac{d}{dt}{\color{SMblack}\alpha_R}=-\gamma_N({\color{SMblack}\alpha_R}-
\alpha_N+\alpha_{N+1}-\alpha_{L}), \\
\frac{d}{dt}\alpha_L=-\gamma_N(\alpha_L+
\alpha_N-\alpha_{N+1}-\alpha_{H}), \\
\frac{d}{dt}\alpha_{N+1}=-(\gamma_N+\gamma_{N+1})\alpha_{N+1}+
\gamma_{N+1}\alpha_{N+2}+ \gamma_N({\color{Seba}\alpha_{N}}-\alpha_{H}+\alpha_{L}).
\label{cohr central}
\end{gather}
These equations describe 1D classical random walk. So,
stationary states for arms decoupled from the central node would
be vectors with equal elements, $\alpha_j=\alpha$ for $j=1\ldots
N$ or $j=N+1\ldots 2N$ and arbitrary $\alpha$.  Also, there is a
stationary state localised in two controlling modes, ${\color{SMblack}a_R}$ and
$a_L$, with ${\color{SMblack}\alpha_R}=\alpha_L$ and $\alpha_j=0$, $j=1\ldots 2N$.
Obviously, for the whole structure, the equal distribution of
amplitudes in both arms $\alpha_j=\alpha$ for $j=1\ldots N$ and
$j=N+1\ldots 2N$, and equal amplitudes in the controlling modes,
${\color{SMblack}\alpha_R}=\alpha_L$ is also the stationary state. Excitation of
just one arm and one of the controlling modes with equal
amplitudes (i.e., for example, $\alpha_j=\alpha$ for $j=1\ldots
N$, ${\color{SMblack}\alpha_R}=\alpha$ and ${\color{SMblack}\alpha_R}=0$, $\alpha_j=0$ for
$j=N+1\ldots 2N$ ) is also a stationary state.
By exciting control modes, ${\color{SMblack}a_R}$ and
$a_L$ in certain states, one can make an initial excitation of a
particular mode propagate either to the one arm, or to another, or
to both arms simultaneously (see Fig.~5 {\color{SMblack} in the main text}). Notice, the such a distributing action can be achieved catalytically, since, as it follows from {\color{SMblack}Supplementary} Eq.~(\ref{cohr central}), the coherence of two controlling modes are conserved, ${\color{SMblack}\alpha_R}(t)+\alpha_L(t)={\color{SMblack}\alpha_R}(0)+\alpha_L(0)$, for any time-moment, $t$. {\color{SMblack} In Fig.~5b,} one can see an illustration of the distribution for the two-arm structure shown in Fig.~5a.

\section*{Supplementary Note 5: Double chain and dissipative localisation}

{\color{SMblack} For the sake of generalisation, now we consider two parallel dissipatively coupled chains as shown in {\color{SMblack}Supplementary} Fig.~\ref{fig3-suppl}.
The chain consists of squares, connected  side by sides,} so, the Lindblad operator of $j$-th square is
\begin{equation}
A_{j}=a_{j,+}-a_{j,-}+a_{j-1,+}-a_{j-1,-}. \label{lsquare0}
\end{equation}
We obtain the following set of equations for the coherent amplitudes:
\begin{equation}
\frac{d}{dt}\vec{\alpha}_j=-\gamma{\hat
O}(2\vec{\alpha}_j-\vec{\alpha}_{j-1}-\vec{\alpha}_{j+1}), \label{eqamp}
\end{equation}
where the matrix ${\hat O}$ has elements ${\hat O}_{j,k}=(-1)^{j+k}$, $j,k=1,2$. The vector $\vec{\alpha}_j=[\alpha_{j,+}, \alpha_{j,-}]^T$. Despite being only a slight modification of the simplest {\color{SMblack}DCC}, the doubled chain has a number of drastically different features. First of all, any vector of coherent amplitudes, $\vec{\alpha}_j$, with equal upper ($+$) and lower ($-$) components is the stationary localised state. Then, initial excitation of any lattice site (say, $\alpha_{j,+}=x$)
gives rise to the stationary state consisting of two-site
localised state $\alpha_{j,+}=\alpha_{j,-}=x/2$ plus delocalised
state $\alpha_{k,+}=(-1)^{j-k}x/2N$,
$\alpha_{k,-}=(-1)^{j-k+1}x/2N$, where $N$ is the number of
systems in each chain. It is interesting that the double chain can serve as an analogous
filter. If both the lower and upper chains {\color{SMblack}are} excited, the
stationary result in each site would be half of the sum of the
lower and upper initial amplitudes. Also, localised states are robust. Additional losses on
sites out of the localisation region do not affect the localised
states. However, they do affect the de-localised stationary
states driving them to the vacuum.

Such localisation phenomena can hold also for infinite perfectly
periodic dissipatively coupled photonic lattices. Let us assume Lindblad
operators of the following form
\begin{equation}
A_j=\sum\limits_{k\in\{n_j\}}x_{jk}a_k.
 \label{lind}
\end{equation}
where $\{n_j\}$ denotes a set of modes coupled to the same dissipative
reservoir; $x_{jk}$ are scalar weights describing such a coupling. To avoid trivial localised states, we assume that there are no isolated sets, and for any $\{n_j\}$ there is a set
$\{n_l\}$ such that the intersection, $\{n_j\}\cap\{n_l\}$, $j\neq l$,
is not empty, but unequal to any of $\{n_j\}$. Additionally, for the ideally periodic structures, we assume that any operator, $a_k$, belongs to at least two different sets, and any set transforms to other set by translation along lattice vectors, ${\vec e}_i$.
Obviously, for any localised stationary state we have $A_j\rho^{{\text{\color{SMblack}loc}}}=0$ $\forall j$. From {\color{SMblack}Supplementary} Eq.~(\ref{lind}) it follows that any localised state occupies at least two sites of the
structure. An example of the honeycomb lattice allowing for dissipative localisation is shown in {\color{SMblack}Supplementary} Fig.~\ref{fig4-suppl} and briefly discussed in the main text.

To demonstrate basic features of dissipative localisation, let us consider here a simple example of a square lattice (see insets in {\color{SMblack}Supplementary} Fig.~\ref{sfig1}). Denoting the sites in the upper left corner of each square as $(j,k)$, we obtain the following Lindblad operators for such a lattice:
\begin{equation}
A_{j,k}=a_{j,k}+a_{j+1,k}+a_{j,k+1}+a_{j+1,k+1}.
\label{lsquare}
\end{equation}
The equations for the amplitudes, {\color{SMblack}Supplementary} Eqs.~(\ref{chainS1},\ref{lsquare}), then read:
\begin{eqnarray}
\frac{d}{dt}\alpha_{j,k}=-\gamma_{j,k}\langle
A_{j,k}\rangle-\gamma_{j+1,k+1}\langle A_{j+1,k+1}\rangle, \label{square eq1}
\\
\frac{d}{dt}\alpha_{j+1,k}=-\gamma_{j,k}\langle
A_{j,k}\rangle-\gamma_{j+1,k-1}\langle A_{j+1,k-1}\rangle.
\label{square eq}
\end{eqnarray}
As {\color{SMblack}can be} seen from {\color{SMblack}Supplementary} Eqs.~(\ref{square eq1}, \ref{square eq}), the
minimal localised states for an infinite square lattice of {\color{SMblack}Supplementary}
Fig.~\ref{sfig1} involve at least four sites (for example, the localised state can be in the set $\{m\}=\{(j+1,k), (j+2,k), (j+1,k+1),
(j+2,k+1)\}$). An example of the localised state composed of coherent states is 
\begin{eqnarray}
|\Psi^{\text{loc}}\rangle=
|\alpha\rangle_{j+1,k}|-\alpha\rangle_{j+2,k} \nonumber \\ 
|\alpha\rangle_{j+2,k+1} |-\alpha\rangle_{j+1,k+1}
\prod\limits_{j,k\notin\{m\}}|0\rangle_{j,k}.\label{localized1}
\end{eqnarray}
Any closed contour including either 0, 2 or 4 sites of every square can host a localised state. 
{\color{SMblack}A finite lattice can also support localised edge states with even, as well as odd, number of sites.}
For example, the three-site edge state in the upper left corner of the lattice shown in the inset of {\color{SMblack}Supplementary} Fig.~\ref{sfig1}
can have the coherent state with amplitudes $2\alpha$  in $(1,1)$ and states with the amplitudes $-\alpha$ in sites $(2,1)$ and $(1,2)$. 

Localised states of a dissipatively coupled lattice can be arbitrarily extended. A state can propagate through the lattice exciting localised states in several cells.
To illustrate the basic features of such propagation, let us consider the dynamics of just a single unit cell of the square lattice [just one $A_{j,k}$
of {\color{SMblack}Supplementary} Eq.~(\ref{lsquare})]. One has
\begin{equation}
\vec{a}(t\rightarrow\infty)=(1-{\bf S}/4)\vec{a}(0), \label{four}
\end{equation}
where the vector of time-dependent modal amplitudes is
$\vec{a}(t)=[\alpha_{1,1}(t), \alpha_{1,2}(t), \alpha_{2,1}(t),
\alpha_{2,2}(t)]^T$ and $\bf{S}$ is the matrix of units. {\color{SMblack}Supplementary}
Eq.~(\ref{four}) shows that the  final result is an initial state
minus the result of complete symmetrisation of it over the cell.
A similar process occurs for the complete lattice. Symmetrical parts
propagate. Curiously, this process is described by
the  classical two-dimensional random walk. Let us introduce
variables $\lambda_{m,n}(t)=(-1)^{m+n}\langle A_{m,n}(t)\rangle$.
For $\gamma_j\equiv\gamma>0$, $\forall j$. From {\color{SMblack}Supplementary} Eqs.~(\ref{square eq1}, \ref{square eq}) it follows that 
\begin{eqnarray}
 \frac{d}{dt}\lambda_{m,n}=-4\gamma\lambda_{m,n}+\gamma\bigl(\lambda_{m+1,n}+\lambda_{m-1,n}+\lambda_{m,n-1}+\lambda_{m,n+1}\bigr).
\label{term}
\end{eqnarray}
Similar heat-like propagation of coherences was found recently in
dissipatively coupled 1D spin chains \cite{our2015}.  An illustration of the
stationary distribution arising from the initial excitation of
just one mode is given in {\color{SMblack}Supplementary} Fig.~\ref{sfig1}a. In {\color{SMblack}Supplementary} Fig.~\ref{sfig1}b,
a spectrum of the equation matrix for {\color{SMblack}Supplementary} Eqs.~(\ref{square eq1}, \ref{square eq}) is given.
The plateau of zero eigenvalues {\color{SMblack} is} separated from the non-zero
eigenvalues with the gap of $\gamma$. {\color{SMblack}Supplementary} Eq.~(\ref{term}) points also to
the existence of delocalised stationary modes  given by the condition
$\langle A_{m,n}\rangle=(-1)^{m+n}\alpha$.

Despite coupling to neighbour sites, the stationary localised state is completely impervious to additional loss even on sites adjacent to those where the stationary
state is localised. It can be seen even for the simplest example of the single-cell system. Taking the equation matrix  Eq.~(\ref{square eq}) with two sites subjected to additional loss with the rate $\gamma$ as $-\gamma({\bf S} -{\mathrm{diag}}(1,1,0,0))$, one gets $\vec{a}(t\rightarrow\infty)={\bf O}\vec{a}(0)$, where
the only non-zero elements of the matrix ${\bf O}$ are
$O_{3,3}=O_{4,4}=0.5$ and $O_{3,4}=O_{4,3}=-0.5$. Again, the
result is the initial vector minus its symmetrisation, but only
for the modes untouched by the additional loss. A similar effect
holds for a larger lattice. In {\color{SMblack}Supplementary} Fig.~\ref{sfig1}c one can see an example of a localised state in the region free of additional loss
arising from the initial excitation of just one initial mode. The
inset in the {\color{SMblack}Supplementary} Fig.~\ref{sfig1}d shows sites affected by additional individual loss with rate $\gamma$. {\color{SMblack}Supplementary} Fig.~\ref{sfig1}d shows eigenvalues of the equation matrix Eq.~(\ref{square eq1}, \ref{square eq}) for this case. Only two localised states survive for the case, and the gap between the zero plateau and decaying modes are closed; there are
modes with decay rates much less than $\gamma$.

Naturally, the localised stationary state can be entangled. The simplest example of the entangled states for the minimal  localised states of the infinite square lattice of
{\color{SMblack}Supplementary} Fig.~\ref{sfig1}a up to the normalization factor is
\begin{eqnarray}
|\Psi^{\text{loc}}\rangle &=
&|\alpha\rangle_{j+1,k}|-\alpha\rangle_{j+2,k}
|\alpha\rangle_{j+2,k+1}
|-\alpha\rangle_{j+1,k+1} \nonumber \\
&+&|-\alpha\rangle_{j+1,k}|\alpha\rangle_{j+2,k}
|-\alpha\rangle_{j+2,k+1} |\alpha\rangle_{j+1,k+1}
\end{eqnarray}
which for $|\alpha|>0$ is  entangled since an averaging
over any mode included in this equation gives a mixed
state. Up to the normalization factor, the reduced state of any three modes is given by
\begin{eqnarray}
 \rho_3=|\psi_3^+\rangle\langle\psi_3^+|+
|\psi_3^-\rangle\langle\psi_3^-|+ 
e^{-2|\alpha|^2}(|\psi_3^+\rangle\langle\psi_3^-|+
|\psi_3^-\rangle\langle\psi_3^+|),
\end{eqnarray}
where $|\psi_3^{\pm}\rangle=|\pm\alpha\rangle_l|\mp\alpha\rangle_m
|\pm\alpha\rangle_n$, and indexes $l,m,n$ number three modes
remained after averaging over the fourth one.

Another example of the localised state is the state with just a single photon distributed
over several lattice {\color{SMblack}sites,} such as
\begin{equation}
|\Psi^{\text{{\color{SMblack}loc}}}\rangle=\sum\limits_{k\in\{n\}}(-1)^l|1\rangle_k,
\label{locferm}
\end{equation}
where the index $l$ numbers modes along the closed contour
connecting all the sites belonging to the set $\{n\}$ where the
state is localised; the state $|1\rangle_k$ corresponds to the photon in $k$-th
mode and the vacuum in all other modes. The same localisation regions
as for the coherent modal states are possible for both the perfect
and finite lattices. For example, the upper-left corner state of
the lattice {\color{SMblack}presented in Supplementary} Fig.~\ref{sfig1}a is
$|\Psi^{\text{{\color{SMblack}loc}}}\rangle=2|1\rangle_1-|1\rangle_2-|1\rangle_3$. The
states of {\color{SMblack}Supplementary} Eq.~(\ref{locferm}) are entangled. Amount of entanglement
is proportional to the number of systems in the localised state.
For example, the generalised Schmidt number for the state
of {\color{SMblack}Supplementary} Eq.~(\ref{locferm}) is $2(N-1)$, $N$ being the number of sites~\cite{guo}. Notice that existence of the localised state, {\color{SMblack}Supplementary} Eq.~(\ref{locferm}), points to {\color{SMblack} the} possibility of dissipative compacton-like localisation in fermionic lattices, too.

\vspace*{0.5cm}

%
%
%
%

\end{document}